%% file: main.tex
\documentclass[conference]{ieeetrans}
\IEEEoverridecommandlockouts
\usepackage{cite}
\usepackage{amsmath,amssymb,amsfonts}
\usepackage{graphicx}
\usepackage{textcomp}
\usepackage{xcolor}
\usepackage{xspace}
\usepackage{graphics}
\usepackage{epsfig}
\usepackage{enumitem}
\newcommand{\Paragraph} [1] {\smallskip\noindent{\bf #1. }}
\usepackage{booktabs}
\usepackage{multirow}
\usepackage{bbding}
\usepackage{url}
\usepackage{tikz}
\usepackage{filecontents}

\usepackage{graphicx}
\usepackage{balance}  
\usepackage{subfig}
\usepackage{float}
\usepackage{verbatim}
\usepackage{color,xcolor}
\usepackage{array}
\usepackage{bbding}

\usepackage{algorithm}
\usepackage{algpseudocode}

\newcommand{\pt}{ExpTM\xspace}
\newcommand{\ptc}{ExpTM-compaction\xspace}
\newcommand{\ptf}{ExpTM-filter\xspace}

\newcommand{\ptfa}{ExpTM-F\xspace}
\newcommand{\um}{ImpTM\xspace}
\newcommand{\umum}{ImpTM-unified-memory\xspace}
\newcommand{\umzc}{ImpTM-zero-copy\xspace}
\newcommand{\umuma}{ImpTM-UM\xspace}

\newcommand{\hybrid}{HyTM\xspace}

\newcommand{\sysname}{{HyTGraph}\xspace}

\newcommand{\best}[1]{\underline{\textbf{#1}}}
\newcommand{\zyf}[1]{\textcolor{black}{#1}}

\def\BibTeX{{\rm B\kern-.05em{\sc i\kern-.025em b}\kern-.08em
    T\kern-.1667em\lower.7ex\hbox{E}\kern-.125emX}}
\begin{document}

\title{{\sysname: GPU-Accelerated Graph Processing with Hybrid Transfer Management}
}

\author{\IEEEauthorblockN{Qiange Wang$^*$,  Xin Ai$^*$, Yanfeng Zhang\textsuperscript{\Envelope} \thanks{ Qiange and Xin contributed equally. Yanfeng is the corresponding author.}, Jing Chen, Ge Yu}
\IEEEauthorblockA{\textit{School of Computer Science and Engineering}\\\textit{Northeastern University,}
Shenyang, China \\
\{wangqiange,aixin0,chenjing\}@stumail.neu.edu.cn;
\{zhangyf,yuge\}@mail.neu.edu.cn}
}

\maketitle

\begin{abstract}
Processing large graphs with memory-limited GPU needs to resolve issues of host-GPU data transfer, which is a key performance bottleneck. Existing GPU-accelerated graph processing frameworks reduce the data transfers by managing the active subgraph transfer at runtime. 
Some frameworks adopt explicit transfer management approaches based on explicit memory copy with filter or compaction. In contrast, others adopt implicit transfer management approaches based on on-demand access with zero-copy or unified-memory.
Having made intensive analysis, we find that as the active vertices evolve, the performance of the two approaches varies in different workloads. Due to heavy redundant data transfers, high CPU compaction overhead, or low bandwidth utilization, adopting a single approach often results in suboptimal performance. 

In this work, we propose a hybrid transfer management approach to take the merits of both the two approaches at runtime, with an objective to achieve the shortest execution time in each iteration. Based on the hybrid approach, we present \sysname, a GPU-accelerated graph processing framework, which is empowered by a set of effective task scheduling optimizations to improve the performance. Our experimental results on real-world and synthesized graphs demonstrate that \sysname achieves up to 10.27X speedup over existing GPU-accelerated graph processing systems including Grus, Subway, and EMOGI.
\end{abstract}


\input{sec1}
\input{sec2}
\input{sec3}
\input{sec4}
\input{sec5}

\input{sec6}

\section*{Acknowledgment}
\vspace{-0.05in}
  We thank the anonymous reviewers for their constructive comments and suggestions. The work is supported by the National Natural Science Foundation of China under grants 62072082, U1811261, the National Social Science Foundation of China under grants 21\&ZD124, and the Key R\&D Program of Liaoning Province 2020JH2/10100037.

\bibliographystyle{abbrv}
\bibliography{main}

\end{document}

%% file: sec1.tex
\vspace{-0.05in}
\section{Introduction}
Analyzing large-scale graph data plays an important role in real-world applications, including geo-information mining, social network analysis, and business association analysis. 

Compared with the shared-memory-based frameworks and the shared-nothing-based frameworks,
GPU-based graph processing attracts more attention for its high memory bandwidth and massive parallelism \cite{medusa_tpds_2014,cusha_hpdc_2014,gunrock_ppopp_2016,tigr_asplos_2018,sep-graph_ppopp_2019}. Unfortunately, GPU's limited device memory can only accommodate a small set of real-world graphs. When the size of the input graph exceeds the GPU memory capacity (\emph{memory oversubscription}), existing GPU-based systems fail to work (e.g., Medusa \cite{medusa_tpds_2014}, CuSha \cite{cusha_hpdc_2014}, Gunrock \cite{gunrock_ppopp_2016}, Tigr \cite{tigr_asplos_2018}, SEP-Graph \cite{sep-graph_ppopp_2019}, etc).

Recently, researches\cite{subway_eurosys_2020,graphreduce_sc_2015,scaph_atc_2020,totem_pact_2012,garaph_atc_2017,ASCETIC_ICPP_2021,emogi_vldb_2020,grus_acmtrans_2021,halo_vldb_2020} focus on supporting GPU-accelerated graph processing to take advantage of both the high-performance GPU graph processing and the large host memory for storing the large-scale graphs. Similar to that out-of-core graph processing \cite{graphchi_osdi_2012,x-stream_sigops_2013,gridgraph_atc_2015,lumos_atc_2019}, the major challenge for GPU-accelerated graph processing is the low computation resource utilization caused by the extensive data movement overhead between GPU and host memory. 
Compared to the high-speed global memory access bandwidth in GPU, the host memory and GPU are connected with a slow PCIe interface, which can be an order of magnitude slower. For example, the host-GPU bandwidth via PCIe 3.0 can be limited to be 16GB/s ({12.3GB/s} in practice) \cite{emogi_vldb_2020}. 
{Moreover, the development of new generation PCIe has not narrowed the bandwidth gap. Even though the PCIe bandwidth has been improved in past years (from 16GB/s to 64GB/s), the bandwidth gap between GPU memory and PCIe is still very large as shown in Table \ref{tab:gpugen}.} 

\begin{table}[t]
	\vspace{-0.05in}\footnotesize
        \centering
    \caption{Advances from NVIDIA P100 to H100.}	
    \label{tab:gpugen}
	\centering
	{\renewcommand{\arraystretch}{1.2}
	\begin{tabular}{ l l l c}
		\hline
		
		\hline
		
		\multirow{2}*{\textbf{GPUs}} &
		\multirow{2}*{\textbf{\footnotesize Mem. bdw.}} &
		\multirow{2}*{\textbf{\footnotesize PCIe x16 bdw.}}& {\textbf{\footnotesize   Mem. bdw/}}\\
		&&&{\textbf{\footnotesize   PCIe. bdw}}\\
		\hline
		{\footnotesize P100 \cite{P100} (2016)} & {\footnotesize 732GB/s} &  {\footnotesize 16GB/s (Gen3) }&\textbf{45.8X} \\
		{\footnotesize V100 \cite{V100} (2017)} & {\footnotesize 900GB/s} &  {\footnotesize 16GB/s (Gen3) }&\textbf{50X} \\
    	{\footnotesize A100 \cite{A100} (2020)}& {\footnotesize 1.9TB/s } & {\footnotesize 32GB/s (Gen4) }&\textbf{48.6X} \\
	{\footnotesize H100 \cite{H100} (2022)}& {\footnotesize 3TB/s} & {\footnotesize 64GB/s (Gen5) }&\textbf{48X} \\
		\hline
		
		\hline
		\vspace{-0.35in}
	\end{tabular}
	}
\end{table}

To reduce the data movements between GPU and host memory, the existing GPU-accelerated frameworks \cite{graphreduce_sc_2015,graphie_pact_2017,subway_eurosys_2020,scaph_atc_2020,emogi_vldb_2020,halo_vldb_2020,grus_acmtrans_2021} track the evolving active vertices during the iterative processing.
Considering a vertex-centric graph processing, where the computation is performed in a series of iterations, in each iteration, \zyf{the algorithm takes only the vertices updated by the previous iteration as input (i.e., active vertices), updates their out-going neighbors and marks the neighbors whose values have been updated as the active vertices in the next iteration}. During the iterative processing, only the out-going edges of the active vertex (i.e., active edges) need to be accessed.
\zyf{Following the existing frameworks\cite{subway_eurosys_2020,emogi_vldb_2020,scaph_atc_2020,graphreduce_sc_2015,graphie_pact_2017,halo_vldb_2020,grus_acmtrans_2021}, we assume that the vertex-associated data (including vertex value, neighbor index, and activity status) can be resident in the GPU memory and the edge-associated data (including edges and edge weights) can entirely fit into the host memory.} During the iterative processing, the active subgraph containing active edges must be transferred to the GPU memory.

\textcolor{black}{According to the way} of reducing host-GPU data transfers, the existing frameworks can be classified into two categories: 
\textbf{\underline{Exp}licit} (active subgraph) \textbf{\underline{T}ransfer \underline{M}anagement (\pt)} based frameworks\cite{graphie_pact_2017,graphreduce_sc_2015,scaph_atc_2020,subway_eurosys_2020,ASCETIC_ICPP_2021} and \textbf{\underline{Imp}licit} ({active subgraph}) \textbf{\underline{T}ransfer \underline{M}anagement (\um)} based frameworks \cite{halo_vldb_2020,grus_acmtrans_2021,emogi_vldb_2020}.

With the {\pt} approach, the programmers have to manually manage the active subgraph transfer. In {\pt}-based frameworks, the oversized graph is partitioned into smaller subgraphs that can fit into GPU device memory. Before being transferred to GPU \textcolor{black}{through explicit memory copy engine} (\texttt{cudaMemcpy}), the subgraphs have to pass through a CPU-based redundancy removal module to remove inactive edges. According to the working mode, this approach can be either filter-based \cite{graphreduce_sc_2015,graphie_pact_2017,gts_sigmod_2016} or compaction-based \cite{scaph_atc_2020,subway_eurosys_2020,ASCETIC_ICPP_2021}, and the transfer reduction performance is determined by the power of removal module.

Recently, a more general solution, {{\um}-based approaches} have become available
\cite{emogi_vldb_2020,grus_acmtrans_2021,halo_vldb_2020}. 
Rather than explicitly managing the data movements of active subgraphs. \zyf{\um-based frameworks allow GPU programs to access the active edges in the host memory in an on-demand mechanism\cite{page_asplos_2015,manager_micro_2017,halo_vldb_2020,emogi_vldb_2020}.} Using {\um} requires little engineering \textcolor{black}{ work}, we can directly extend the single GPU frameworks into an out-of-core one by managing the host-resident edge data with unified-memory \cite{halo_vldb_2020,grus_acmtrans_2021} or zero-copy memory \cite{emogi_vldb_2020}. Then during the iterative processing, the memory unit containing active edges can be implicitly transferred to the GPU memory without programmers' manual management. Since \um approaches rely on the system-provided memory access mechanism, its transfer efficiency is sensitive to the graph access pattern. Recent research \cite{emogi_vldb_2020} shows that the performance gap can reach more than three times.

\begin{table}[t]
	\vspace{-0.05in}\small
	\caption{Runtime comparison of Subway and EMOGI on variable algorithms and datasets.}
	\label{tab:intro}
	\centering
	\vspace{-0.05in}
	{\renewcommand{\arraystretch}{1.2}
	\begin{tabular}{c | c c| c c}
		\hline
		
		\hline
		&\multicolumn{2}{c|}{SK-2005 graph}&\multicolumn{2}{c}{PageRank Algorithm}\\
		\cline{2-5}
		&{ \textbf{SSSP}} &
		{ \textbf{PageRank}} &
		{\textbf{ sk-2005 }}& {\textbf{ uk-2007}}\\
		\hline
		Subway&14.6(s) & \underline{\textbf{8.7}(s)} &  \underline{\textbf{8.7}(s)}&16.9(s) \\
		EMOGI&\underline{\textbf{7.5}(s)} & 18.6(s) &  18.6(s)&\underline{\textbf{12.4}(s) }\\
		\hline
		
		\hline
	\end{tabular}
	}
\end{table}

Having made extensive analysis, \zyf{
we find that a decision to choose one or the other approach for the best performance is determined by the memory access pattern of active edges. In a GPU-accelerated graph processing framework based on a single approach, the performance is often suboptimal.} We show the performance comparison of Subway\cite{subway_eurosys_2020} (a \ptc-based framework) and EMOGI\cite{emogi_vldb_2020} (an \umzc-based framework). Table \ref{tab:intro} shows that on sk-2005 graph\cite{sk_UK}, EMOGI outperforms the Subway on \textcolor{black}{Single Source Shortest Path algorithm (SSSP)}
, but it \textcolor{black}{losses} on PageRank. In contrast, for PageRank algorithm, Subway beats EMOGI on SK dataset\cite{sk_UK}, but \textcolor{black}{losses} on UK dataset\cite{sk_UK}.

In this paper, we present a \textbf{\underline{Hy}brid \underline{T}ransfer \underline{M}anagement} approach (\textbf{\hybrid}). Unlike prior frameworks that use either \textbf{\pt} or \textbf{\um}, our hybrid approach combines \pt and \um to maximize the performance. In the preprocessing stage, \hybrid partitions the graph as \pt does. Then during the iterative processing, it estimates \pt cost and \um cost on-the-fly by analyzing the edge access pattern of each partition and chooses the most cost-efficient transfer approach. 
Based on the \textbf{\hybrid}, we propose \sysname, a GPU-accelerated graph processing system with flexible asynchronous task scheduling. Unlike prior frameworks\cite{subway_eurosys_2020,scaph_atc_2020,graphreduce_sc_2015,graphie_pact_2017} that simply process the loaded subgraphs multiple times, \sysname adopts \textcolor{black}{a contribution-driven priority scheduling method.} 
Through a lightweight \textcolor{black}{graph reorganization}, \sysname can gather and prioritize the vertices that have a large distribution to convergence.

We have made the following contributions in this paper.

\begin{itemize}[leftmargin=*]

\item Providing insights into the two existing approaches. We conduct a comprehensive study on the performance merits and limits of the two transfer management approaches (\pt and \um).

\item Proposing a hybrid transfer management framework. We introduce a hybrid transfer management method to maximize the performance by taking the merit of both \pt and \um.

\item Delivering a GPU-accelerated graph processing system. Based on the hybrid transfer management method, we design and implement \sysname, a transfer-efficient GPU-accelerated graph processing system with flexible asynchronous task scheduling to enable high performance. 

\end{itemize}
We evaluate \sysname on both real-world and synthesized graphs. The experimental results show that \sysname outperforms the state-of-the-art systems, i.e., on average 4.61X speedup over Subway\cite{subway_eurosys_2020}, 2.37X speedup over Grus\cite{grus_acmtrans_2021} and 1.74X speedup over EMOGI\cite{emogi_vldb_2020}.

%% file: sec2.tex
\begin{figure}
	\centering
	\vspace{-0.1in}
	\includegraphics[width=0.46\textwidth]{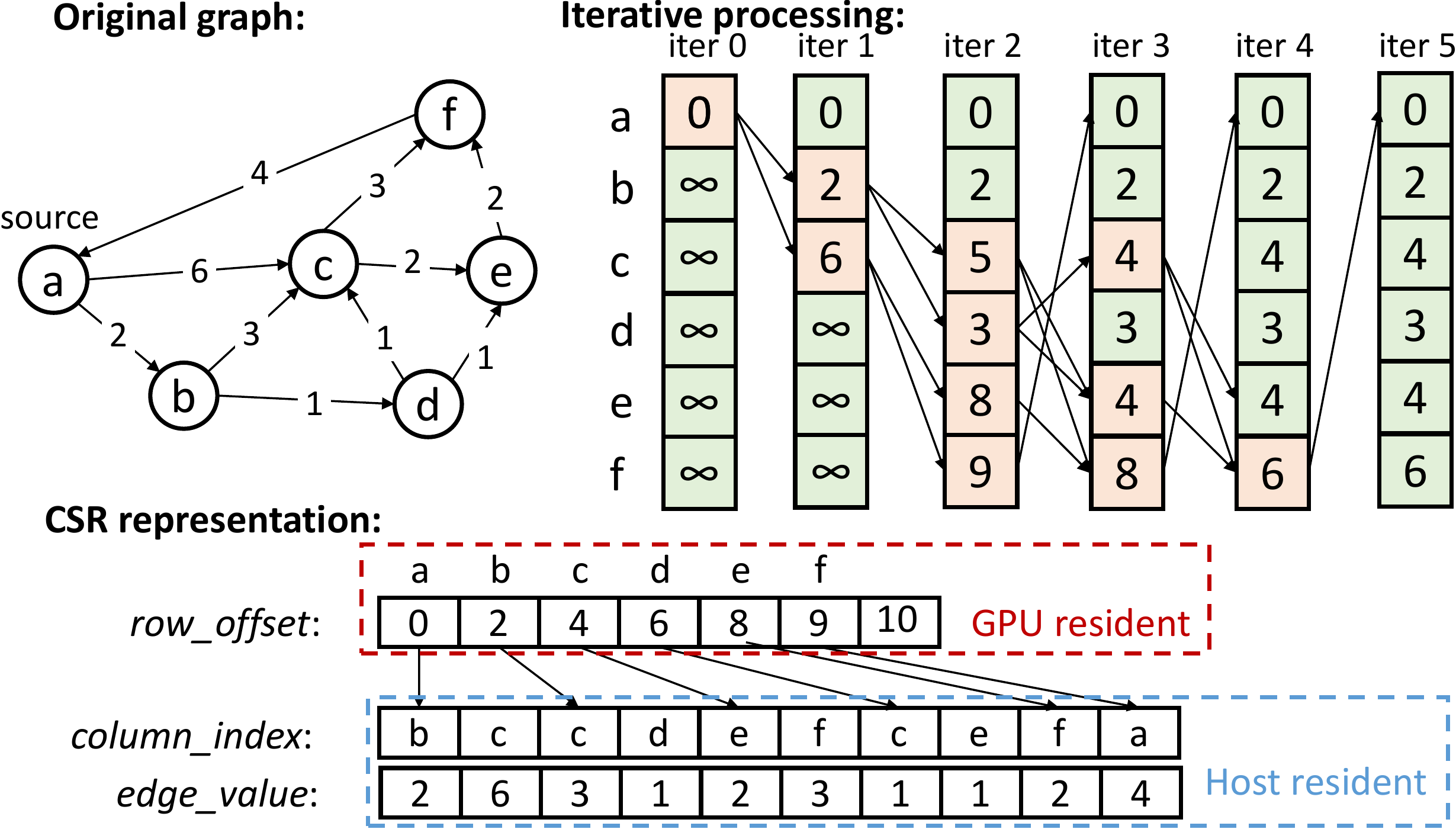}
	\caption{An example \textcolor{black}{of vertex-centric SSSP computation starting from source \texttt{a}.} The orange box represents the active vertex and the green box represents the inactive vertex. The input graph is organized into CSR, whose vertex-associated data is resident in the GPU and the edge-associated data is resident in the host memory.}
	\label{fig:CSR}
	\vspace{-0.15in}
\end{figure}

\section{Background}
\vspace{-0.03in}
\label{sec:background}

\subsection{Vertex-Centric Graph Processing and Active vertices}

{Vertex-centric programming} \cite{pregel_sigmod_2010,powergraph_osdi_2012} has been widely adopted in Graph processing frameworks, for its simplicity, high scalability, and powerful expression ability. It defines a generic function that defines the behavior of a vertex and its neighbors. Considering the message passing direction, the function can be either pull-based or push-based\cite{sep-graph_ppopp_2019}. During the computation, this function is evaluated on all input vertices iteratively until the algorithm reaches convergence. Figure \ref{fig:CSR} illustrates a push-based example of {SSSP}, an algorithm to find the shortest paths from a given source vertex to all the other vertices. It starts from the source vertex $a$, where the initial distance is set to $0$. In each iteration, each input vertex sends its current shortest distance to the outgoing neighbors, and the neighbor receiving messages will update its shortest distance as the shortest one. When no vertices {is updated}, the algorithm converges. We can observe that, during the iterative computation, only the vertices updated by the previous iteration (active vertices) need to be processed and its number increases with the message scatter from the source vertex and decreases as most vertices converge.

{
The graph processing which processes graph data iteratively has a special memory access pattern. The edge data that requires substantial memory footprint is read-only, and the vertex data that requires small memory footprint is read-write. In a GPU-accelerated platform where the input graph exceeds the GPU memory capacity, placing the relatively small vertex data in GPU and accessing the required edge data on demand from host memory is a worth trying approach. Firstly, The edge data transfer is easier to manage than the vertex data transfer because the edge data is read-only, requiring only one-way communications (host-to-GPU). Secondly, in real-world graphs, the number of vertex is often orders-of-magnitude less than the number of edge. Even a commonly used 16GB GPU can still process a large graph with hundred-millions of vertices and tens of billions of edges. As this design needs to retransfer the edge-associated data multiple times, adopting additional transfer management module to reduce the inactive edge transfers is critical to performance.
}

 \begin{figure*}
	\centering
	\includegraphics[width=0.95\textwidth]{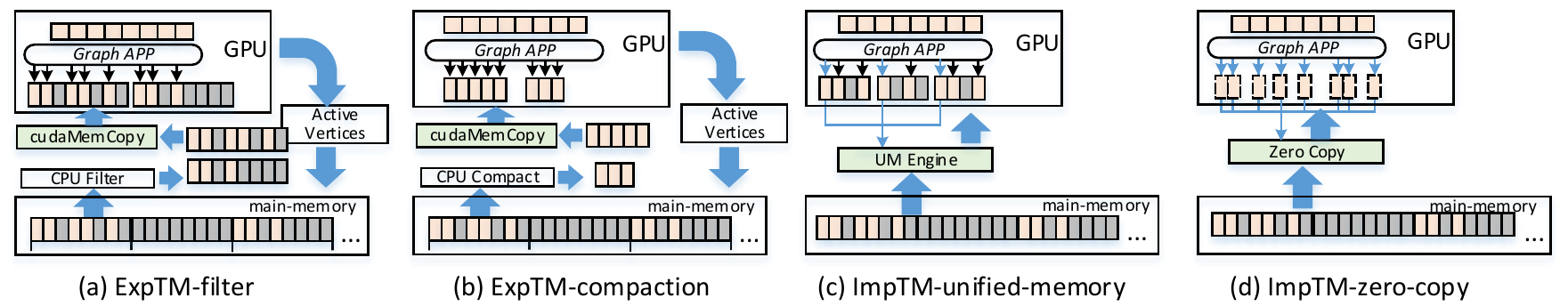}
	\vspace{-0.08in}
	\caption{An example of the four approaches, \pt-based approaches need to transfer the active vertices back to the host side for removing the inactive edges, \um-based methods does not need this process. The thin blue arrow, thin black arrow, and thick blue arrow represent the remote memory access, local memory access, and host-GPU data transfer, respectively.}

	\label{fig:example_comparison}
	\vspace{-0.05in}
\end{figure*}

\subsection{\textbf{ExpTM} Approaches}
\Paragraph{\zyf{\ptf}} GraphReduce\cite{graphreduce_sc_2015}, GTS\cite{gts_sigmod_2016}, and Graphie\cite{graphie_pact_2017} adopt filter-based method to reduce the inactive subgraph transfer. They monitor the active edges of the partitioned subgraphs in each iteration and {transfers} 
only those partitions that contain active edges. Figure \ref{fig:example_comparison} (a) provides an illustrative example. 
This method only filters out partitions that do not contain active edges without doing additional processing, so the entire partition will be directly transferred to GPU even if only one edge is active. \zyf{ When the proportion of active edges in a partition is low, the volume of redundant data transfer will be large.}

\Paragraph{\ptc} In contrast, some frameworks \cite{scaph_atc_2020,ASCETIC_ICPP_2021,subway_eurosys_2020} introduce CPU-assisted compaction to reduce redundant data transfers. Before transferring a partitioned subgraph to GPU, these frameworks use CPUs to remove the inactive edges and compact the remaining edges into a continuous memory space to leverage explicit memory copy. Figure \ref{fig:example_comparison} (b) shows an illustrative example. 
Subway \cite{subway_eurosys_2020} is a typical \ptc-based system.
In each iteration, it compacts all the active edges into a new graph and transfer it to GPU for parallel processing.
Compared with the filter-based frameworks \cite{graphie_pact_2017,graphreduce_sc_2015}, compaction-based frameworks can minimize the data transfers by removing all inactive edges. But at the cost, it involves heavy CPU and main memory read/write overhead.

\subsection{\textbf{ImpTM} Approaches}

\Paragraph{\umum}
The unified-memory defines a managed memory space in which both GPU and CPU can observe a single address space with a coherent memory image \cite{halo_vldb_2020,grus_acmtrans_2021}. The memory pages ({4KB in default}) containing the requested data are automatically migrated to the devices, and the subsequent accesses to the same page will read data from the GPU memory without additional data transfers.
When the memory footprint of the kernel is larger than the GPU memory, some pages may need to be evicted from the GPU to make room for the new pages. Figure \ref{fig:example_comparison} (c) shows an illustrative example. 
Notice that the ``automated migration'' cost is not free. When a requested memory page is not in the GPU memory, a page fault is triggered, which requires not only data transfer but also heavy \zyf{Translation Lookaside Buffer (TLB) invalidation and page table updating overhead \cite{emogi_vldb_2020}.}

\Paragraph{\umzc}
In contrast, zero-copy memory access is a more lightweight approach. Zero-copy maps pinned host memory to GPU address spaces, allowing GPU programs to directly access the host memory through the Transaction Layer Packet (TLP) of PCIe\cite{emogi_vldb_2020}. Compared with unified-memory, zero-copy provides much fine-grained access granularity. By the PCIe 3.0 specification, each TLP can process at most 256 \zyf{outstanding} memory requests simultaneously, and each request can carry 32/64/96/128-byte\cite{emogi_vldb_2020} data \textcolor{black}{according to} the size of accessed data. Therefore, zero-copy allows the programs to access the edges of multiple randomly distributed active vertices simultaneously, and each vertex occupies one or several memory requests. In addition, zero-copy requires less transferring overhead than unified-memory based frameworks because it requires no additional page migration overhead. As a sacrifice, the zero-copy method cannot provide the data reuse function. Multiple accesses to the same data will cause multiple separate data transfers.

%% file: sec3.tex
\begin{figure*}
	\centering
	\includegraphics[width=0.95\textwidth]{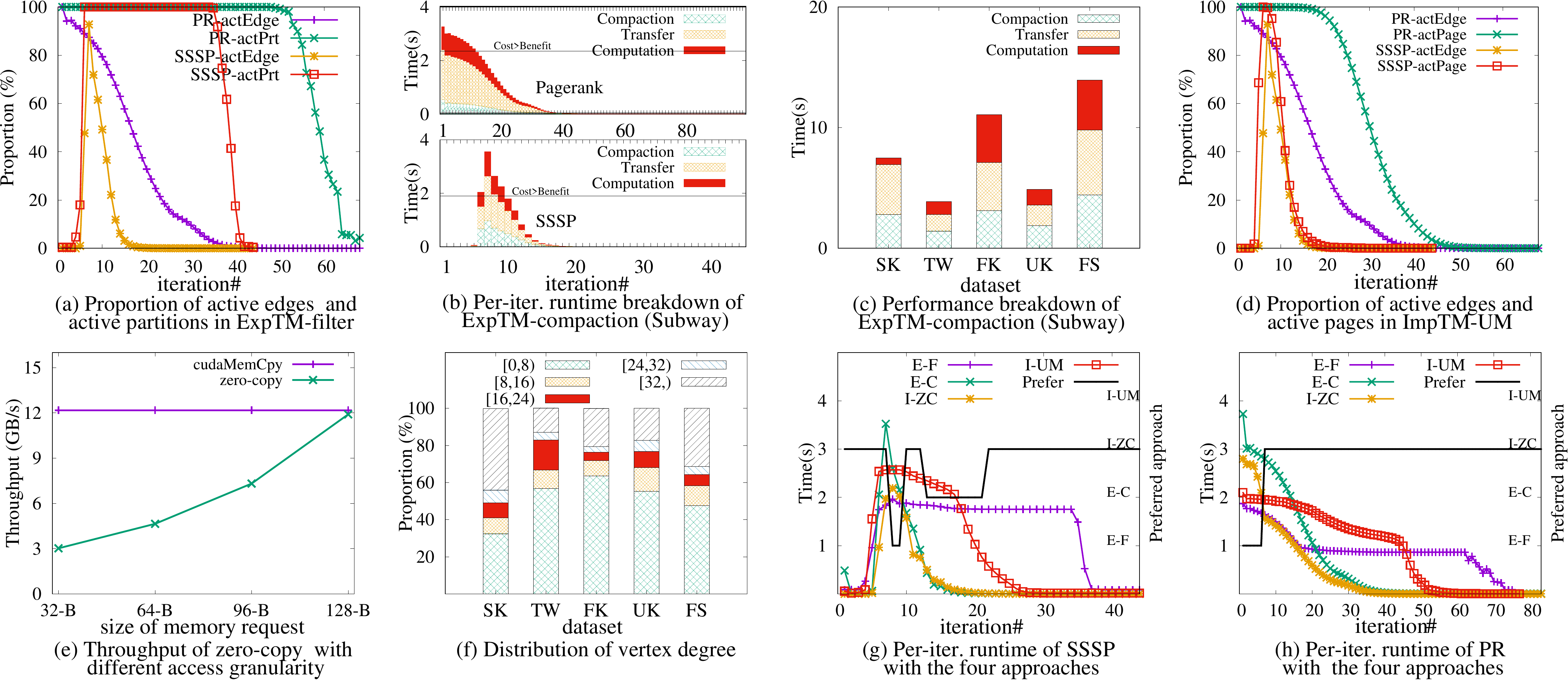}
    
	\caption{\zyf{Performance analysis of four engine of the two approaches.}}
	\label{fig:motivation}
	\vspace{-0.18in}
\end{figure*}
\section{ Analysis of existing approaches: A Motivating Study}
\label{sec:analysis}
In this section, we experimentally analyze the existing approaches with two graph algorithms SSSP and PageRank, because they have two typical active vertices change patterns (increase then decrease, and monotone decrease). The details of the used graphs, test platform, and system configurations are given in Section \ref{sec:expr:setup}.  

{
\subsection{Analysis of \pt}
\vspace{-0.05in}
\Paragraph{\ptf}
As mentioned above, the filter-based \pt has a large volume of redundant transfers even if the proportion of active edge is low. We run PageRank and SSSP on friendster-konect \cite{FK_tw} graph to explore the redundant data transfer problem, the partition number is set to 256.
Figure \ref{fig:motivation} (a) shows the proportion curves of active edges and partitions containing active \textcolor{black}{edges} (active partitions). \zyf{We can observe that the proportion of active partitions does not decrease immediately with the proportion of active edges.}
For SSSP and PageRank algorithms, the active edges account for only 28.3\% and 12.3\% of the total transfer volume. Therefore, \pt-filter is inefficient when there are few active edges in the partition. While, when the proportion of active edge is large, \pt-filter method shows advantages, because it can fully utilize the PCIe bandwidth with {\texttt{cudaMemcpy}}.

\Paragraph{\ptc}
The compaction-based \pt achieves significant transfer reduction and can leverage the efficient explicit memory copy. But it involves heavy additional {active edge compaction overhead}, which is positively correlated to the proportion of active edges. 
As pointed out by Subway\cite{subway_eurosys_2020}, when the proportion of active edges is large (e.g., 80\%), the cost of compaction can even outweigh the benefit of transfer reduction\cite{subway_eurosys_2020}, {Figure \ref{fig:motivation} (b)} illustrates the per-iteration runtime breakdown of Subway (a \pt-compaction based framework) and 
indicates \zyf{when the costs outweigh the benefits}.
{Figure \ref{fig:motivation} (c)} illustrates the overall performance breakdown of SSSP algorithm on Subway, {we} remove its preprocessing stage and show only the execution time. We can observe that on all five datasets, {the compaction stage accounts for 34.5\% of the overall runtime.} 

\vspace{-0.05in}
\subsection{Analysis of \um}
\vspace{-0.05in}
\Paragraph{\umum}
\zyf{
Unified-memory}
is not an efficient way \textcolor{black}{of} 
handling graph algorithms. First, the cost of ``automated migration'' is high, due to heavy TLB invalidation overhead and page table updating overhead \cite{emogi_vldb_2020}, the peak bandwidth of unified-memory can only reach 73.9\% of that of explicit memory copy (\texttt{cudaMemcpy})\cite{emogi_vldb_2020}. Second, the graph algorithms usually have poor {temporal} locality\cite{emogi_vldb_2020,ASCETIC_ICPP_2021}. When accessing the vertex with only several or dozens of neighbors, the 4KB memory page may still contain non-negligible inactive data  \cite{emogi_vldb_2020,halo_vldb_2020}. Figure \ref{fig:motivation} (d) shows the proportion of the active edges and the active memory pages of each iteration, for SSSP and PageRank algorithms, the active edges account for only 54.5\% and 65.0\% of the total transfer volume. 
{For these two reasons, the Unified-memory-based \umzc shows poor transfer efficiency on large graphs, no matter the proportion of active edge is high or low. However, the UM-based method will have good performance when the graph size is small enough to fit into GPU memory because the graph can be fully cached in the GPU memory after being transferred once.}

\Paragraph{\umzc} The key challenge of implementing efficient zero-copy based graph processing is \zyf{fully utilizing the PCIE bandwidth.} As pointed out by EMOGI\cite{emogi_vldb_2020}, saturating most of the 256 memory requests in each TLP with 128-byte data is necessary for maximizing the PCIe bandwidth utilization. In addition to the payloads of memory requests, the TLP also includes a header field to maintain the necessary control information. A smaller memory request size means that PCIe needs to use more TLPs to process the same amount of data, and thus wastes more bandwidth on transferring the header fields. {Figure \ref{fig:motivation} (e)} shows the 
throughput of zero-copy under different memory request granularity (from 32 byte to 128 byte). We can observe that, when the memory request size is 128-byte, the zero-copy access can achieve almost the same performance as \texttt{cudaMemcpy} (the maximum PCIe utilization). While, when the access granularity is set to 32-byte, the throughput decreases significantly. To achieve the maximum bandwidth utilization, EMOGI \cite{emogi_vldb_2020} proposes merged and aligned access with which each warp of threads access consecutive neighbors of one vertex in a 128-byte cache line size from the edge-associated array. In this way, the neighbors of high-degree vertices can be accessed with consecutive and saturated memory requests.
However, guaranteeing most of the memory requests reach 128-byte is challenging. Assuming each vertex occupies 4-byte, we need 32 neighbors per vertex to saturate the 128-byte memory requests. In real-world graphs, the number of neighbor is often less than this value due to the power-law property. \zyf{Figure \ref{fig:motivation} (f)} illustrates the distribution of vertex degrees of five real-world graphs used in this paper. \zyf{Most vertices (on average 74.7\%) have less than 32 neighbors, and 51.1\% of them have less than 8 neighbors.}
Zero-copy based method has unstable performances on real world graphs, it prefers subgraphs with few active vertices and large average degrees.

}
\begin{table*}[!t]
	\caption{Summary of existing systems}
	\hspace{-0.05in}
	\label{tab:system_overview}
	\centering
	{\renewcommand{\arraystretch}{1.2}
	\begin{tabular}{l l l l l}
		\hline
		
		\hline
		\textbf{Approach}&{\textbf{Systems}} &
		{\textbf{Strengths}} &
		{\textbf{Weaknesses}}&{\textbf{Prefer}}\\
		\hline
		\multirow{3}*{\ptf}&{GraphReduce}\cite{graphreduce_sc_2015} &$\bullet$Less CPU overhead &$\bullet$ Redundant data&$\bullet$Subgraph with a large\\
	    &{Graphie}\cite{graphie_pact_2017} &$\bullet$High transfer efficiency& transfers&proportion of  active edges\\
		&{GTS}\cite{gts_sigmod_2016} &&&\\
		\hline
		\multirow{3}*{\ptc}&{Subway}\cite{subway_eurosys_2020}&$\bullet$Significant transfer&$\bullet$ High compaction& $\bullet$\zyf{Subgraph} with a small \\
		&{Scaph}\cite{scaph_atc_2020}& reduction&overhead& proportion of active edges\\
	    &{Ascetic}\cite{ASCETIC_ICPP_2021}& & & and small average degree\\
		\hline
		\multirow{2}*{\umum}&{HALO}\cite{halo_vldb_2020} &$\bullet$ Easy to use &$\bullet$ Redundant data transfers& $\bullet$ Small graph that can  \\
		&{Grus}\cite{grus_acmtrans_2021} & &$\bullet$ High transfer overhead& fit into GPU memory \\
        \hline
		\multirow{3}*{\umzc}&\multirow{3}*{EMOGI\cite{emogi_vldb_2020}} &$\bullet$ Easy to use&$\bullet$ Unstable bandwidth &$\bullet$ \zyf{Subgraph} with a small\\
		&&$\bullet$ Fine grained memory &utilization&proportion of active edges\\
		&&access&&and high average degree\\
		
		\hline
		
		\hline
		\vspace{-0.15in}
	\end{tabular}
	}
	\vspace{-0.1in}
\end{table*}
\subsection{Performance Comparison of the Four Approaches}
\label{sec:analysis:compare}

We report the per-iteration runtime of \ptf, \ptc, \umum, and \umzc on friendster-konect\cite{FK_tw} with two typical graph algorithms (the traversal algorithm SSSP and the iterative algorithm PageRank\cite{sep-graph_ppopp_2019}) in Figure \ref{fig:motivation} (f) and (g). We implement \ptf (E-F), \umum (I-UM), and \umzc (I-ZC) with SEP-Graph's processing kernel\cite{sep-graph_ppopp_2019}. Since the GPU does not write data back to host memory, we open the \texttt{cudaMemAdviseSetReadMostly} optimization for \umum (the evicted memory pages will be discarded instead of 
written back to host memory). We use Subway \cite{subway_eurosys_2020} as the \ptc (E-C), because of its highly-optimized CPU compaction engine and kernel function from Tigr\cite{tigr_asplos_2018}.
All the approaches are configured with synchronous processing to ensure that the number of active vertices in each iteration is roughly the same. 

\begin{figure}
\vspace{-0.1in}
	\centering
	\includegraphics[width=0.5\textwidth]{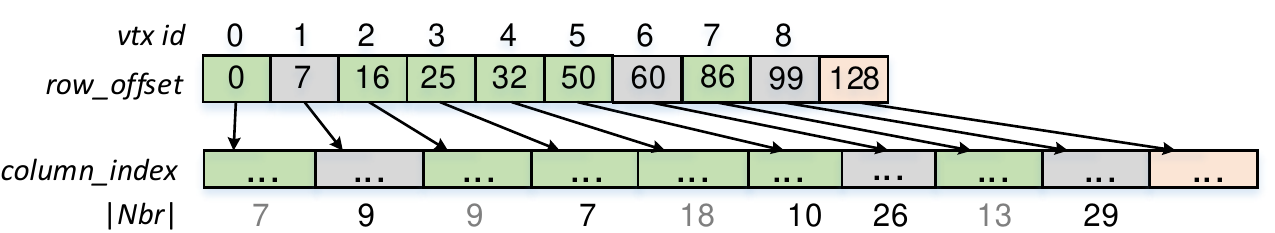}
	\vspace{-0.1in}
	\caption{A toy graph with 9 vertices and 128 edges in CSR. The graph is divided into two subset, each of which containing 64 edges. The numbers below are the number of neighbors.}
	\label{fig:active_degree}
	\vspace{-0.13in}
\end{figure}

We use a ``\textbf{Prefer}'' curve to indicate the winner in each iteration.
By referring to the proportion curves of active edges of SSSP and PageRank in Figure \ref{fig:motivation} (a), we can observe that when the proportion of active edges is large, \ptf has better performance because it has high bandwidth utilization (with \texttt{cudaMemcpy}) and requires no additional CPU processing overhead. When the proportion of active edge is small, \umzc shows better performance than the others in most iterations because it can transfer the neighbors needed by the active vertices with fine-grained memory requests. For SSSP algorithm, \ptc shows better performance than \umzc on some iterations. 
This can be attributed to the unstable performance of zero-copy under different vertex degrees. As mentioned above, the performance of zero-copy is not only related to the proportion of active edges, but also related to the number of active vertices. When the number of active edges is fixed, a large number of active vertices means that zero-copy has to use more unsaturated memory requests to process the data and thus results in more TLPs. Figure \ref{fig:active_degree} shows a toy graph with 9 vertices and 128 edges. We divide the graph into two subsets (in green and gray), each of which has 64 neighbors.  The two subgraphs have the same proportion of active edges (0.5) when being activated. When the subgraph with 6 vertices (in green) is activated, zero-copy has to use 6 memory requests. When the subgraph with 3 vertices is activated, zero-copy only needs 3 memory requests. This causes zero-copy performance to be unstable, even if their proportions of active edge are the same. Therefore, neither \ptc nor \umzc shows consistently better performance than each other. 

In summary, although existing approaches significantly reduce the data transfers, the performance is still suboptimal. Most of them can only adapt to one or several cases.

\subsection{Summary of Existing Systems}
In Table \ref{tab:system_overview}, We summarize these approaches and their representative systems. We also list their strengths, weaknesses, and preferred subgraph. In addition to the systems \cite{graphreduce_sc_2015,subway_eurosys_2020,halo_vldb_2020,emogi_vldb_2020} mentioned above, Scaph \cite{scaph_atc_2020} and Ascetic \cite{ASCETIC_ICPP_2021} adopt \ptc. Different from Subway, Scaph \cite{scaph_atc_2020} performs compaction on the partitioned graph. It distinguishes the partitions with a small proportion of active edges, and compacts them for the subsequent GPU processing. In contrast, the partitions with a large proportion of active edges will be entirely loaded to GPU. Ascetic\cite{ASCETIC_ICPP_2021} divides GPU memory into a static region and an on-demand region, exploits the temporal locality across iterations for the static region, and compacts the other active subgraphs with CPU for the on-demand region.
Grus\cite{grus_acmtrans_2021} is an \um-based framework. It manages the edge-associated data in main memory with priorities, prefetching high-priority data to the GPU through unified-memory and accessing low-priority data through zero-copy. In addition, some frameworks \cite{totem_pact_2012,garaph_atc_2017} also use CPU-GPU collaboration to accelerate graph processing. We will review these works in Section \ref{sec:related}. 

%% file: sec4.tex
\begin{figure}
	\centering
	\includegraphics[width=1\linewidth]{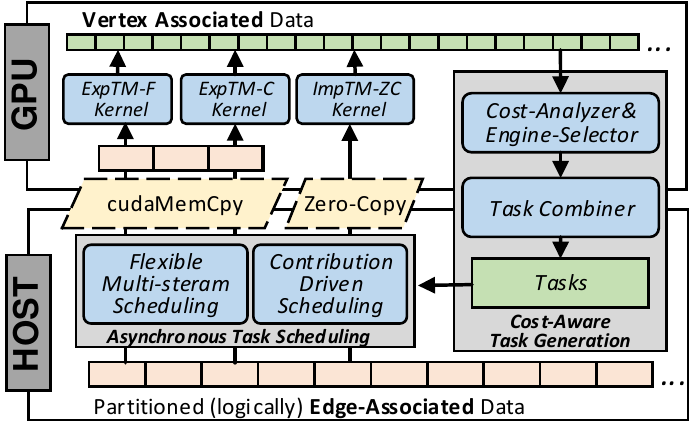}
	\caption{Overview of \sysname.}
	\label{fig:workflow}	
	\vspace{-0.1in}
\end{figure}

\section{\sysname Overview}
We present \sysname, a GPU-accelerated graph processing framework that adopts hybrid transfer management (\hybrid) to maximize performance. \sysname organizes the graph into CSR structure, whose neighbor index array is resident in the GPU global memory, and edge-associated arrays (neighbor array and edge-weight array) are logically partitioned on the host side. Following the existing frameworks\cite{scaph_atc_2020,graphie_pact_2017}, \sysname partitions the edge-associated data into $N$ edge-balanced partitions $\{P_0,P_1\ldots,P_{N-1}\}$ with chunk-based partitioning\cite{gemini_osdi_2016,scaph_atc_2020}, in which each $P_i$ is a set of consecutively numbered vertices of partition $i$.
During the iterative computation, the partitions containing active edges are scheduled with their most cost-efficient engine for the GPU computation.
\sysname provides two functions to achieve efficient \hybrid.

\Paragraph{Cost-aware task generation} In the cost-aware task generation module, \sysname computes the data transfer costs of different approaches and selects the most cost-efficient engine for each partition. Based on the analysis in section \ref{sec:analysis}, we choose \ptf, \ptc, and \umzc as our baseline. In addition, \sysname provides a task combiner to merge the subgraphs (to be scheduled) into larger tasks to achieve lower scheduling overhead in the task scheduling stage.

\Paragraph{{{Asynchronous task scheduling}}}
\sysname introduces asynchrony to improve task scheduling efficiency. Rather than simply recompute the loaded subgraph multiple times\cite{subway_eurosys_2020,scaph_atc_2020}, \sysname adopts a contribution-driven priority scheduling to prioritize those partitions that contribute more to convergence. This method is based on the following observation: those vertices with large degrees often become the hub in the computation paths. To improve resource utilization, \sysname uses multiple CUDA streams to overlap the computation kernel, data transfer, and CPU-based active subgraph compaction.

Figure \ref{fig:workflow} shows an overview of \sysname. The cost-aware task generation and asynchronous task scheduling are iteratively alternating until the algorithm reaches convergence.

\section{Cost-Aware Task Generation}

\subsection{Cost Analysis and Engine Selection}
\label{sec:hybrid:check}

Most of the Existing activeness-tracking-based frameworks use the activeness ratio as the metric \cite{graphreduce_sc_2015,graphie_pact_2017,garaph_atc_2017,scaph_atc_2020}. They evaluate the proportion of active edges on each partitioned subgraph to determine the appropriate processing engine. Such an approach provides an intuitive and lightweight distinguishing method, but is hard to adapt to \hybrid approach. As discussed in Section \ref{sec:analysis:compare}, the proportion of active edge cannot reflect the time cost of different approaches. In this work, we present a cost-aware processing engine selection method.
During the iterative processing, we measure the overhead for each partition as follows.

\Paragraph{Cost of \ptf}
The \ptf based approach entirely transfers the partitions with active edge entirely to GPU device memory with explicit memory copy engine (\texttt{cudaMemcpy}). So it has only data transfer cost, which can be approximated by the saturated TLPs (as discussed in Section \ref{sec:analysis}, Figure \ref{fig:motivation} (e)). Given a partition $i$, the number of memory transaction can be calculated with $\sum_{v\in P_i}{D_o(v)}*d_1/m$, where $\sum_{v\in P_i}{D_o(v)}$ is the number of edge of partition $i$, $d_1$ represents the memory occupation of one vertex, and $m$ represents the maximum capacity 
of an outstanding memory request (128-byte). Denote $MR$ as the maximum number of an outstanding memory request in TLP ($MR=256$ in PCIe 3.0 specification) and $\lceil\cdot\rceil$ as the round-up operation, we formalize the transfer overhead of each partition $i$ as follow:

\vspace{-0.03in}
\small
\begin{align}
\label{eq:cost_ptf}
Tef_i=\Big\lceil \big (\sum_{v\in P_i}{D_o(v)}\big)*d_1/m/MR\Big\rceil*RTT,
\end{align}
\normalsize
\vspace{-0.03in}

\zyf{where $\Big\lceil \big (\sum_{v\in P_i}{D_o(v)}\big)*d1/m/MR\Big\rceil$ is actually the number of TLPs, and RTT represents the round trip time for PCIe to process each saturated TLP.}

\Paragraph{Cost of \ptc} \ptc involves additional CPU-based compaction, so its cost consists of two parts, the data transfer overhead, and the compaction overhead. Since the compaction needs to reorganize the active edges and change their positions, we also need to generate a vertex index array and transfer it to GPU for addressing the compacted neighbors. Then the transfer volume can be formalized as $\sum_{v\in A_i}{D_o(v)}*d_1+|A_i|*d_2$, where $A_i$ represents the active vertex subset of $P_i$ and $d_2$ represents the memory occupation of each index. The CPU-based compaction is related to transfer volume and the throughput of CPU-based compaction, which can be computed with $\sum_{v\in A_i}{D_o(v)}*d_1+|A_i|*d_2/Thpt_{cpt}$, where $Thpt_{cpt}$ is the throughput of CPU-based compaction. Then the cost of \ptc can be formalized as follow:

\vspace{-0.03in}
\small
\begin{align}
\label{eq:cost_tec}
Tec_i=\Big\lceil \big (\sum_{v\in A_i}{D_o(v)}*d_1+|A_i|*d_2\big)/m/MR\Big\rceil*RTT\notag\\
+\sum_{v\in A_i}{D_o(v)}*d_1+|A_i|*d_2/Thpt_{cpt}
\end{align}
\normalsize
\vspace{-0.03in}

\Paragraph{Cost of \umzc} The \umzc approach provides vertex-oriented on-demand access in a cacheline size, so each active vertex $v$ takes one or several independent memory requests. The memory request number of vertex $v$ can be formalized as $\lceil D_o(v)*d_1/m\rceil$. $D_o(v)$ represents the number of out-going neighbors of active vertex $v$. Considering that we can hardly guarantee the neighbors of all vertices start from the aligned memory position, some vertices may have the misaligned neighbor array and thus require one additional memory transaction\cite{emogi_vldb_2020}. We introduce a function $am()$, which returns 1 for the vertices requiring one additional transaction and 0 for the others\footnote{In the implementation, the memory request number of each active vertex $\lceil D_o(v)*d_1/m\rceil+am(v)$ can be directly computed by using the length and physical start position of the neighbors.}. Then the transfer overhead of partition $i$ can be formalized as follow:

\vspace{-0.03in}
\small
\begin{align}
\label{eq:cost_tie}
Tiz_i= \Big\lceil\Big(\sum_{v\in A_i}\big(\lceil D_o(v)*d_1/m\rceil+am(v)\big)\Big)/MR\Big\rceil*RTT_{zc},
\vspace{-0.1in}
\end{align}
\normalsize
\vspace{-0.03in}

where $\Big(\sum_{v\in P_i(V)\cap A_i}\lceil D_o(v)*d_1/m\rceil+am(v)\big)$ is the required memory transactions of active vertices.
It should be noted that
the TLP round trip time of zero-copy ($RTT_{zc}$) is not the same as that in \pt ($RTT$) because the payload of each TLP in zero-copy may be unsaturated. This makes $RTT_{zc}$ always less than the $RTT$s in \ptf and \ptc. In this paper, we use a dumpling factor $\gamma$ to compute $RTT_{zc}$ for each partition as follows: $RTT_{zc}=\gamma*RTT+(1-\gamma)*(\sum_{v\in A_i}{D_o(v)}/\sum_{v\in P_i}{D_o(v)})*RTT$, where $(\sum_{v\in A_i}{D_o(v)}/\sum_{v\in P_i}{D_o(v)}$ is the proportion of active edge. $\gamma*RTT$ represents the fixed time to process a TLP, and $(1-\gamma)*(\sum_{v\in A_i}{D_o(v)}/\sum_{v\in P_i}{D_o(v)})*RTT$ represents the time related to the size of payload. By referring to \cite{emogi_vldb_2020}, we set $\gamma$ to $0.625$.

\Paragraph{Transfer engine selection} 
We need to compare $Tef_i$, $Tec_i$, and $Tiz_i$ to choose the most cost-efficient transfer engine. While theoretically modeling the throughput of compaction operation $Thpt_{cpt}$ in $Tec_i$ (formula \ref{eq:cost_tec}) is challenging because \ptc introduces parallel and random writes on the main memory. This makes $Thpt_{cpt}$ vary with active edges nonlinearly. In practice, we compute $Tec_i$ by considering only the transfer overhead and compare it with $Tef_i$ and $Tiz_i$.

If $Tec_i$ is less than $ \alpha*Tef_i$ and $Tec_i $ is less than $\beta*Tiz_i$, we choose \ptc. The first condition comes from Subway's observation \cite{subway_eurosys_2020}, where $\alpha$ is set to $80\%$. The second condition is based on the following observation from Section \ref{sec:analysis}: When a partitioned subgraph has few active edges but many active vertices, the average degree of these active vertices is small, and zero-copy requires multiple unsaturated memory requests to transfer the data. Therefore, compacting and transferring them with \ptc is a better choice. In our implementation, $\beta$ is set to $40\%$.
If these conditions are not met, we compare $Tiz_i$ and $Tef_i$. If $Tiz_i$ is less than $Tef_i$, we choose \umzc. Otherwise we choose \ptf. \zyf{In the computation, the value of RTT can be arbitrarily specified, because it will be omitted during comparison}.

Since the cost computation between partitions is independent, \sysname computes $Tef_i$, $Tec_i$, and $Tiz_i$ and chooses the most cost-efficient transfer engine on GPU, transferring only the selection result back for the subsequent processing. This design can help reduce the burden of CPUs. 
\zyf{We show the overall execution flow of the cost-aware engine selection in algorithm \ref{alg:htm}, line (2-13)}.

\begin{algorithm}[t]
	\caption{Cost-aware task generation} \small 
	\label{alg:htm}  
	\begin{algorithmic}[1]
		\Require active vertex set $\{A_0,\cdots, A_{N-1}\}$ of $N$ partitions,
		\Ensure tasks prefer \pt-filter $\{ Vf_{0}\ldots Vf_{M-1}\}$ ($M<N$), task prefer \pt-compaction $Vc$, and task prefer \um-zero-copy $Vz$.
		\State initialize a selection array $\{p_0,\ldots p_{N-1}\}$ on GPU.
		\newline \underline{ \textbf{Cost analysis and engine selection:}}
		\For{each $A_{i}$ in $\{A_0,\cdots, A_{N-1}\}$} in parallel
		\State Compute $Tef_i$, $Tec_i$, and $Tiz_i$ according to Formula (\ref{eq:cost_ptf},\ref{eq:cost_tec},\ref{eq:cost_tie})
		\If{$Tec_i<\alpha*Tef_i$ and $Tec_i<\beta*Tiz_i$}
		\State $p_i$=`\pt-C';
	    \State insert $A_i$ to $Vc$; \textcolor{gray}{//pre-combine on GPU}
		\ElsIf{$Tef_i<Tiz_i$}
		\State $p_i$=`\pt-F';
		\Else
		\State $p_i$=`\um-ZC';
		\State insert $A_i$ to $Vz$; \textcolor{gray}{//pre-combine on GPU}
		\EndIf
		\EndFor
		\State {\texttt{Copy}} $Vc$, $\{p_0,\ldots p_{N-1}\}$ and $\{A_0,\cdots, A_{N-1}\}$ to host.
		\newline \underline{ \textbf{Task Combination:}}
		\State $i = 0$, $j = 0$, $length = 0$;
		\While{$i<N$}
	    \If{$p_i$==`\pt-F' and $length < k$}
	    \State insert $A_i$ to $Vf_j$;
	    \State $length = length + 1$;
	    \Else
	    \State $length = 0, j = j + 1$;
	    \EndIf
	    \State $i = i + 1$;
		\EndWhile
	\end{algorithmic}  
\end{algorithm}

\subsection{Task Combination}
\vspace{-0.03in}
\label{sec:sys:combine}
Another key to implementing hybrid transfer management is to determine appropriate task scheduling granularity. The existing frameworks\cite{scaph_atc_2020,graphie_pact_2017,graphreduce_sc_2015,garaph_atc_2017} directly use the partitioned subgraphs as scheduling unit. This method is simple but may lead to low efficiency in the task scheduling stage. If the partition size is too large, the coarse-grained cost computation may lead to inappropriate engine selection and thus affect the overall performance. In contrast, if using a small partition size, the transfer engine can be finely selected, but a large number of partitioned subgraphs may cause non-negligible scheduling overhead  (e.g., kernel launches and fragmented data transfers) in the execution stage. Especially on those partitions with few active vertices, even a partition with one active vertex still needs to launch one CUDA kernel. 

To achieve fine-grained engine selection and low overhead task scheduling simultaneously, \sysname decouples the graph partitioning and task partitioning to optimize them separately. \sysname partitions the graph into small partitions (32MB each partition) to provide fine-grained cost analysis. While in the iterative processing, \sysname packages the partitions choosing the same engine into large task units to reduce the scheduling overhead. Specifically, for partitions using \ptf, \sysname merges $k$ consecutive partitions into a large one ($k$=4 in \sysname) to reduce the processing overhead (Line 15-24 in algorithm \ref{alg:htm}). For partitions using \ptc, \sysname merges all their active vertices and writes their neighbor to one consecutive memory space to leverage efficient explicit memory copy (line 6 in algorithm \ref{alg:htm}). For partitions using \umzc, \sysname merges all their active vertices (line 11 in algorithm \ref{alg:htm}) and processes them with one CUDA kernel to leverage the implicit transfer-computation overlapping of zero-copy. 

\section{\zyf{Asynchronous Task Scheduling}}
\vspace{-0.03in}
\label{sec:sys:scheduling}

\sysname improves the asynchronous task scheduling from two directions: First, it accelerates convergence and reduces transfer volume through contribution-driven priority scheduling. Second, it improves resource utilization through multi-stream scheduling.

\subsection{Contribution-Driven Priority Scheduling}
\vspace{-0.05in}
\label{sec:sys:hub}
Asynchronous computation allows the newly updated results to be used immediately in subsequent computation, which has been proved to be effective in GPU-based graph processing\cite {groute_sigplan_2017,sep-graph_ppopp_2019}. \zyf{Many GPU-accelerated graph processing frameworks\cite{graphie_pact_2017,subway_eurosys_2020,scaph_atc_2020} also adopt asynchronous processing to reduce the host-GPU data transfers.}
In these frameworks, the subgraphs loaded to GPU memory will be processed multiple times to squeeze all possible updates in each data transfer. However, simply processing the transferred subgraph multi-times may lead to inefficiency because these local updates may be abolished by the subsequent results from other partitions, leading to more additional computations and data transfers. This problems is known as stale computation problem\cite{aap_sigmod_2018,powerlog_sigmod_2020}. In the real-world experiment, we observe that the multi-round computation can even increase the transfer volume (See Section \ref{sec:expr:perf:trans} for details) in some cases. To effectively leverage the flexibility of asynchronous processing, \sysname adopts contribution-driven priority scheduling.

\Paragraph{Hub-vertex-driven priority scheduling}
Due to the power-law property of real-world graphs, some important vertices with high incoming/outgoing degrees often become the hubs in the computation path. These vertices become critical upstream dependencies of a large number of vertices because of the large outgoing degree. On the other hand, because of the large incoming degree, these vertices have a high probability of being activated in the iterative computation. If these vertices do not accumulate sufficient effective updates before being scheduled, the downstream computation results based on the current value are likely to be abolished by subsequent new updates. Based on this observation, we propose a hub-vertex-driven priority scheduling approach. By ensuring that the hub vertices accumulate enough contributions before being scheduled, \sysname can reduce the possible stale computations on the downstream vertices.
Implementing hub-vertex-driven scheduling in GPU-accelerated platforms is challenging, because the hub vertices may distribute randomly among the whole graph, which makes host-GPU hub-vertex scheduling hard to design. To solve this problem, \sysname adopts the hub sorting method\cite{hubsorting_BIGDATA_2017} to gather and sort the top 8\% important vertices at the beginning of the CSR structure, where the importance score of each vertex $v$ is measured by the following formula:

\small
\vspace{-0.1in}
\begin{align}
    H(v)= \frac{D_o(v)*D_i(v)}{D_{omax}*D_{imax}}
\end{align}\normalsize

$D_i(v) $, $D_o(v)$, $D_{imax}$, and $D_{omax}$ represent the incoming-, outgoing-, maximum incoming-, and maximum outgoing- degree, respectively. In this way, the hub vertices are gathered together, and the non-hub-vertices remain their natural order. \sysname recomputes the loaded subgraph only once because most updates can only pass two hops effectively \cite{lumos_atc_2019}. Another benefit of this hub-vertex gathered method is that the vertices having a high probability of being activated (with large in-degree) are stored together. This property can help improve the effect of cost-aware task generation.

It is worth mentioning that the hub sorting does not need to be performed in each run. As long as performing the hub-sorting once in the data preparation stage, all the subsequent executions (of different algorithms) can benefit from it.

\Paragraph{$\Delta$-driven priority scheduling} For some iterative graph algorithms based on value accumulation, e.g., $\Delta$-based PageRank and PHP algorithm\cite{maiter_corr_2017}, the contribution of vertices is directly reflected in their delta values (the messages to-be-accumulated). Prioritizing the vertices with large $\Delta$ value can help the downstream vertices accumulate updates more effectively \cite{maiter_corr_2017,powerlog_sigmod_2020,sep-graph_ppopp_2019,gunrock_ppopp_2016}. 
Since the original $\Delta$-driven priority scheduling is vertex-centric, it can not be directly used in GPU-accelerated graph processing. To address this problem, \sysname implements $\Delta$-driven scheduling with minor modifications. In each iteration, \sysname computes $\Delta$ value for all partitions and prioritizes those with large delta values. Similar to that of hub-vertex-driven priority scheduling, in $\Delta$-driven scheduling, \sysname process the loaded partition only one more time.

\begin{figure}
\vspace{-0.1in}
	\centering
	\includegraphics[width=0.65\linewidth]{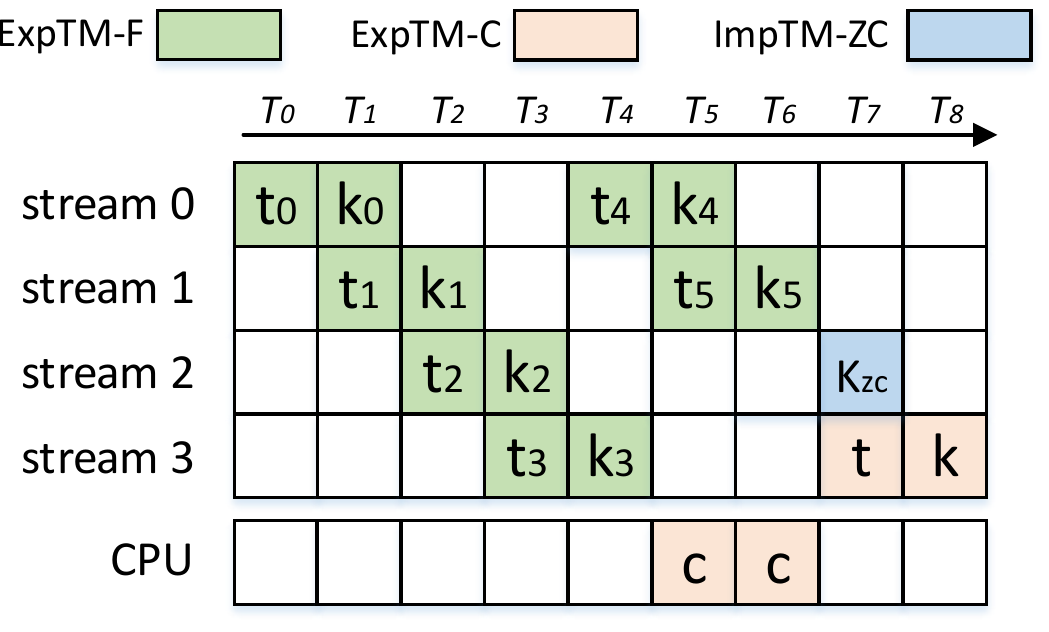}
	\caption{An illustrative example of multi-stream scheduling, the ``t'' represents the transfer operation, ``k'' represents the computation kernel, and ``c'' represents the CPU-based compaction. }
	\label{fig:task}	
	\vspace{-0.1in}
\end{figure}

\subsection{Flexible Multi-Stream Scheduling}
The processing engines of \pt-F, \pt-C, and \umzc-ZC require different resources, including CPUs for active edge compaction, GPU for the computation kernel, and PCIe for the host-GPU data transfer. To overlap the resource utilization and improve the parallelism,
\sysname uses multiple CUDA streams to process the tasks concurrently. Figure \ref{fig:task} shows an illustrative example. During the iterative processing, the task scheduler monitors the available streams and assigns them to tasks that have not been scheduled. The operating system will automatically overlap data transfer and kernel computation of different streams. \sysname first schedules the \pt-Filter tasks with specific priority (as discussed in Section \ref{sec:sys:hub}) to leverage the contribution-driven priority scheduling. 
Then the \umzc and \ptc tasks are scheduled. The CPU-based active edge compaction can be overlapped with the kernel computation and data transfer of \umzc and \ptf.
After finishing all the computing tasks, \sysname will call Algorithm \ref{alg:htm} to prepare for the next iteration.

\subsection{Other Implementations}
\label{sec:sys:optim}
\Paragraph{Implementation of processing kernels}
\sysname provides three processing kernels for implementing \ptf, \ptc, and \umzc hybrid execution. Since the \pt-based engine needs to perform computation on partitioned subgraphs, we implement the processing kernels of \pt by extending SEP-Graph's processing kernel to enable neighbor shifting on the edges-associated array \cite{sep-graph_ppopp_2019}. While for \umzc, \sysname uses the original kernel of SEP-Graph. \sysname inherits a series of inner-GPU optimizations from SEP-Graph, including data-/topology-driven switching\cite{sep-graph_ppopp_2019} and Cooperative Thread Array (CTA) scheduling\cite{CTA_HPCA_2014}. In addition, we implement bitmap-directed frontier optimization\cite{grus_acmtrans_2021} to reduce the atomic conflict of active vertex maintenance.

\Paragraph{Implementation of compaction}
We implement a simple yet efficient parallel edge compaction engine by referring to Subway\cite{subway_eurosys_2020}. Since the physical locations of the edge-associated data are changed in the compaction stage, \sysname has to generate a new compressed neighbor index array and transfers it to the GPU along with the compacted edge array(s) for the \ptc computation.

%% file: sec5.tex
\section{Experimental Evaluation}
\label{sec:expr}
\subsection{Experimental Setup}
\label{sec:expr:setup}
\Paragraph{Environments}
Our test platform is equipped with one Intel Silver 4210 2.20Ghz 10-core CPU, 128GB DRAM, and a NVIDIA GTX 2080Ti GPU with 34SMX clusters, 4352 cores, and 11GB GDDR6 global memory. The GPU is enabled with CUDA 10.1 runtime and 418.67 driver, the host side is running Ubuntu 18.04 with Linux kernel version 4.13.0. All the source codes are compiled with \texttt{O3} optimization.

\begin{table}[!t]
	\vspace{-0.1in}
	\caption{Dataset description.}
	\label{tab:Dataset}
	\centering
	{\renewcommand{\arraystretch}{1.2}
	\begin{tabular}{ l r r c c c}
		\hline
		
		\hline
		
		{\textbf{Dataset}} &
		{\textbf{$|$V$|$}} &
		{\textbf{$|$E$|$}}  &
		{\textbf{$|$E$|$/$|$V$|$}}&
		{\textbf{Size}}\\
		\hline
		
		{sk-2005}\cite{sk_UK} (SK) & 50.6M & 1.93B &38 &28GB  \\
		{twitter}\cite{FK_tw} (TW) &52.5M& 1.96B &37 & 32GB\\
	    {friendster-konect}\cite{FK_tw} (FK) & 68.3M &2.59B & 37 & 42GB\\
		{uk-2007}\cite{sk_UK} (UK) & 105.1M &3.31B & 31& 55GB\\
		{friendster-snap}\cite{FS} (FS)& 65.6M & 3.61B& 55& 58GB\\

		{RMAT}\cite{RMAT_2004_SIAM}& 1-100M &0.1-6.4B & -&- \\
		\hline
		
		\hline
		\vspace{-0.2in}
	\end{tabular}
	}
\end{table}
\Paragraph{Graph algorithms and datasets}
 We evaluate \sysname with four algorithms. Besides SSSP and PageRank, the other two algorithms are Breadth-First Search (BFS) and Connect Component (CC)\cite{sep-graph_ppopp_2019}.
We use both real-world graphs and synthesized graphs in our evaluation. The major parameters of graph datasets that are used in our experiments are presented in Table \ref{tab:Dataset}: Friendster-konect (FK) and friendster-snap (FS) are undirected social network datasets. Sk-2005 (SK) and uk-2007 (UK) are directed web graph datasets. Twitter (TW) is a directed social network dataset. The synthesized graphs used in our experiment are generated by RMAT\cite{RMAT_2004_SIAM} with the power-law distribution.

\Paragraph{Systems for comparison}
We compare \sysname with three representative and public available GPU-accelerated graph processing systems and a CPU-based graph processing system Galois\cite{galois_sosp_2013} (Scaph\cite{scaph_atc_2020} and Ascetic\cite{ASCETIC_ICPP_2021} are also available but we could not run them in our environment due to various CUDA errors, we were not able to resolve these errors after multiple email exchanges with the authors). Besides Subway \cite{subway_eurosys_2020} and EMOGI\cite{emogi_vldb_2020}, Grus is a hybrid framework\cite{grus_acmtrans_2021} that combines \umum and \umzc, when the storage space is large enough, it caches the transferred data in GPU through unified memory. When the device memory is full, Grus accesses the host data through zero-copy. Unlike \sysname, Grus' hybrid processing does not consider the processing overhead of the two approaches.
In addition to these systems, we also implement pure \ptf and \umum in \sysname's codebase for a fair comparison. We use the default configuration of these systems and all the runtime results are measured by averaging the results of 5 runs.

\begin{table}
\vspace{-0.1in}
\renewcommand{\arraystretch}{1.2}
\centering
	\caption{Comparison with other systems.}
	\label{tab:overall}
\begin{tabular}{|c|c|c|c|c|c|c|} 
\hline
\multicolumn{7}{|c|}{Overall runtime (s)}\\  
\hline
Alg.&System&SK&TW&FK&UK&FS\\ 
\hline
\multirow{6}*{PR}&Galois&21.3&66.3&293.6&28.5&342.4\\
\cline{2-7}
&\pt-F&37.7&34.8&60.7&34.3&162.8\\
\cline{2-7}
&\um-UM&6.89&16.5&75.4&22.4&102.7\\
\cline{2-7}
&Grus&\best{1.72}&12.2&52.2&14.8&79.8\\
\cline{2-7}
&Subway&8.68&38.1&73.7&16.9&108.4\\
\cline{2-7}
&EMOGI&18.6&21.4&51.1&12.4&68.3\\
\cline{2-7}
&\sysname& 2.85 & \best{11.5}& \best{30.1} & \best{4.71} &\best{40.8} \\ 
\hline

\hline
\multirow{6}*{SSSP}&Galois&26.7&12.9&51.5&15.2&33.1\\
\cline{2-7}
&\pt-F&60.9&15.1&50.4&60.9&70.1\\
\cline{2-7}
&\um-UM&12.7&10.1&37.2&18.6&34.9\\
\cline{2-7}
&Grus&25.2&11.2&70.8&5.32&16.9\\
\cline{2-7}
&Subway&14.6 &10.9&20.8 & 18.4 & 27.7 \\
\cline{2-7}
&EMOGI&7.46& 4.09&14.9&4.71 &11.8 \\
\cline{2-7}
&\sysname&\best{6.11}&\best{2.09}&\best{8.81}&\best{2.78}&\best{6.64}\\
\hline

\hline
\multirow{6}*{CC}&Galois&23.9&15.7&35.9&55.1&39.4\\
\cline{2-7}
&\pt-F&21.9&5.47&10.9&41.6&11.8\\
\cline{2-7} 
&\um-UM&\best{1.43}&1.49&3.27&7.88&4.16\\
\cline{2-7}
&Grus&{2.09}&1.36&3.21 & 5.17&4.69\\
\cline{2-7}
&Subway&11.67&6.52&8.61&14.7&14.1\\
\cline{2-7}
&EMOGI&4.01&1.96&2.71&4.54&3.76\\
\cline{2-7}
&\sysname&3.65&\best{1.19}&\best{2.01}&\best{3.86}&\best{2.59}\\
\hline

\hline
\multirow{6}*{BFS}&Galois&16.2&7.55&12.5&15.2&14.7\\
\cline{2-7}
&\pt-F&20.3&3.86&8.87&25.1&9.54\\
\cline{2-7}
&\um-UM&1.13&1.29&1.97&2.33&6.25\\
\cline{2-7}
&Grus&\best{0.83} & 1.11&1.85 &2.37 &3.35\\
\cline{2-7}
&Subway&7.39&5.79&6.85&9.04&13.49\\
\cline{2-7}
&EMOGI&1.06&1.04&\best{1.44}&1.26&\best{1.97}\\
\cline{2-7}
&\sysname&0.93&\best{0.85}&1.82&\best{0.88}&2.54\\
\hline
\end{tabular}
\end{table}

\subsection{Overall Performance}
\label{sec:expr:overall}
\subsubsection{Comparison with \pt-F, Subway, and EMOGI}

Table \ref{tab:overall} shows the overall results. Due to the heavy redundant transfer, \pt-F shows worse performance than the others, the speedup of \sysname over \pt-F ranges from 2.01X (for PageRank on FK) to 28.52X (for BFS on UK) with an average of 8.99X. Neither Subway nor EMOGI is always better than the other. The speedup of \sysname over Subway ranges from 2.36X (for SSSP on FK) to 10.27X (for BFS on UK) with an average of 4.61X. Subway's critical performance bottleneck lies in its heavy CPU-based compaction and preprocessing (For SSSP algorithm, the preprocessing and compaction overhead account for 46.9\%-74.9\% of the total runtime). On CC, SSSP, and PageRank, \sysname is faster than EMOGI by 1.96X on average, with its speedups ranging from 1.10X to 6.53X. With the help of zero-copy, EMOGI achieves significant performance improvement on low-activeness subgraphs. While for the high-activeness subgraphs, especially those with dense and small degree vertices, EMOGI usually has low host-GPU utilization due to unsaturated memory requests. In contrast, \sysname achieves efficient data transfer on both high-activeness and low-activeness partitions by adopting hybrid transfer management.
On BFS, \sysname outperforms Subway and EMOGI on SK, TW, and UK. On FK and FS, EMOGI shows better performance because most of the accesses on these two graphs are sparse. Moreover, compared with \sysname, EMOGI avoids the cost analysis, engine selection, and task merging overheads.

\subsubsection{Comparison with Unified-Memory-based Approaches (\umuma and Grus)}

On SK graph, the unified-memory-based frameworks show better performance than the others for PageRank, CC, and BFS algorithms because the edge-associated data can be entirely cached in the GPU memory. UM-based approaches only transfer the data once. While, when processing large graphs, the performance of \umuma degrades significantly because the implicit data transfer requires expensive page replacement and data transfer overhead. The experimental results show that on the four large graphs, \sysname achieves on average 13.14X and 2.37X speedups over \um-UM and Grus, respectively.

{
\subsubsection{Comparison with CPU-based Approach}
From Table \ref{tab:overall}, we can observe that the GPU-accelerated graph processing frameworks show significant performance improvement over CPU-based Galois. Specifically, \sysname shows on average 5.27x-12.78x speedups over Galois. 
}

\subsection{Execution Path Analysis}
\label{sec:expr:perf:exec_path}
\begin{figure}
\vspace{-0.1in}
	\centering
	\includegraphics[width=0.48\textwidth]{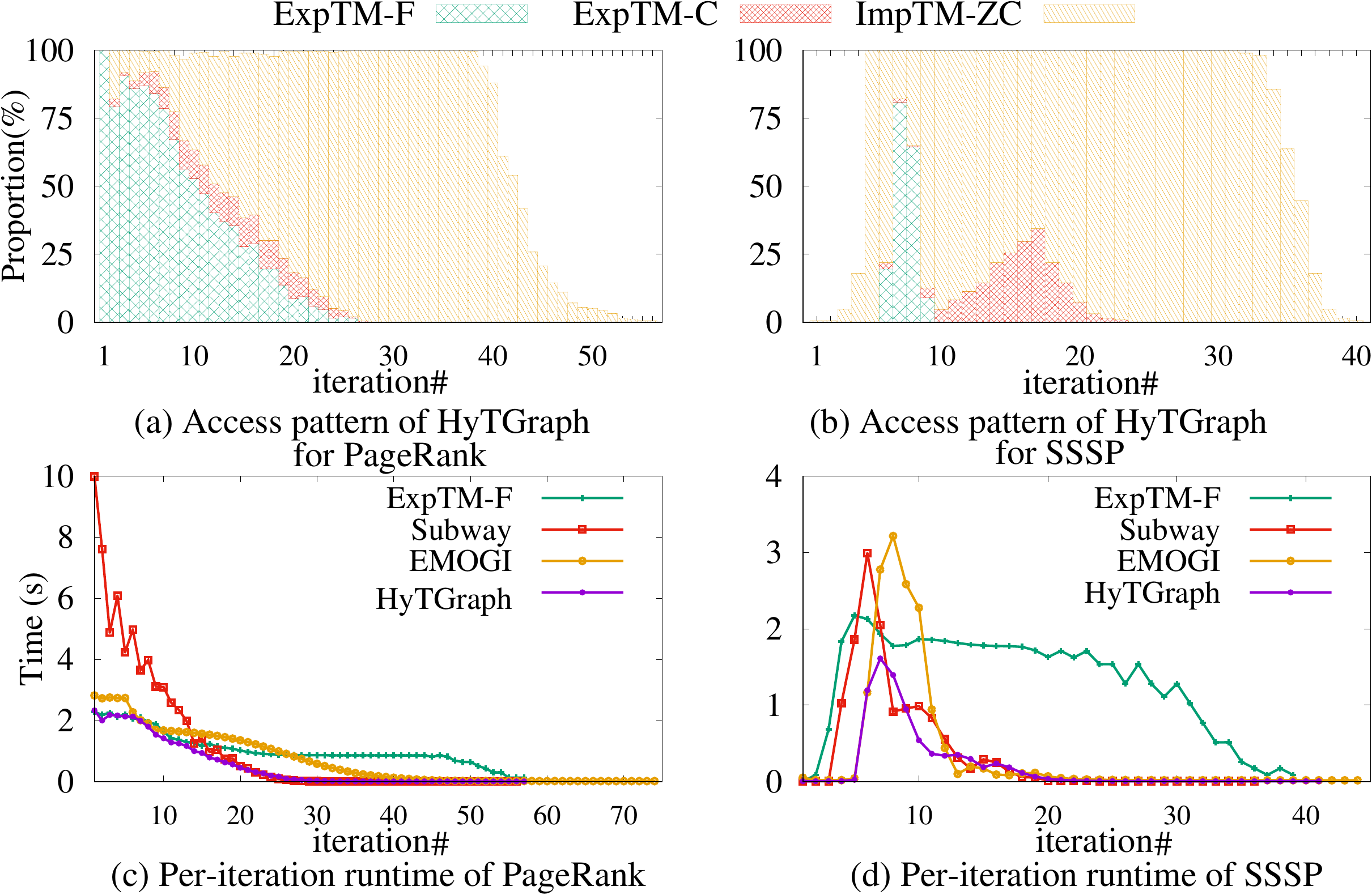}
	\caption{Execution path of \sysname and per-iteration runtime comparison with \pt-filter, EMOGI and Subway (FK).}
	\label{fig:exec}
	\vspace{-0.1in}
\end{figure}

\begin{figure*}[h]
	\centering
	\vspace{-0.1in}
	\includegraphics[width=0.85\textwidth]{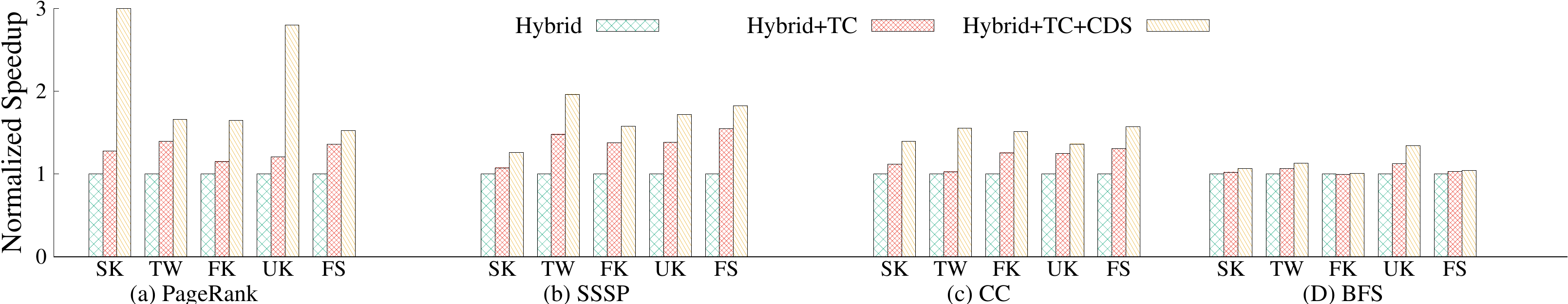}
	\caption{Performance gain analysis of \textbf{T}ask \textbf{C}ombining (TC) and \textbf{C}ontribution-\textbf{D}riven \textbf{S}cheduling (CDS).}\vspace{-0.05in}
	\label{fig:gain}
	\vspace{-0.15in}
\end{figure*}

To demonstrate the performance improvement of hybrid processing, we record the execution path of \sysname on PageRank and SSSP to show the proportion of partitions using \ptf, \ptc, and \umzc in each iteration. Figure \ref{fig:exec} (a) shows the result of PageRank, the proportion of active partitions is high in the early iteration, and \sysname prefers \ptf. As the algorithm converges and many vertices become inactive, the proportion of \umzc begins to increase. For SSSP in Figure \ref{fig:exec} (b), there are few active vertices in the early and last few iterations, and \sysname prefers \umzc. When most vertices are activated in the middle iterations, \sysname prefers \ptf to improve the transfer efficiency. As the number of active vertex decreases, \ptc is also used on some partitions.

{
Figure \ref{fig:exec} (c) and (d) show the per-iteration runtime results of \ptfa, Subway, EMOGI, and \sysname. 
As these systems adopt different asynchronous processing strategies, the active vertex number of different systems in each iteration is not exactly the same. \sysname cannot consistently outperform the others in each iteration. While, through the hybrid transfer management, \sysname can achieve the minimum overall runtime.

}

\subsection{Transfer Reduction Analysis}
\label{sec:expr:perf:trans}
\begin{table}[h]
\vspace{-0.1in}
\renewcommand{\arraystretch}{1.2}
\centering
\caption{Transfer reduction analysis.}
	\label{tab:reduction}
\begin{tabular}{|c|c|c|c|c|c|} 
\hline
\multicolumn{6}{|c|}{Transfer volume / Edge volume}\\  
\hline
Alg.&Dataset&\pt-F&Subway&EMOGI&\sysname\\ 
\hline
\multirow{5}*{PR}&SK&57.6X& 2.46X & 3.31X& \best{2.17X}\\
\cline{2-6}
&TW&52.4X&  \best{5.48X} & 20.6X & 10.9X\\ 
\cline{2-6}  
&FK&58.3X&   \best{10.74}X & 24.6X& 12.01X\\
\cline{2-6}
&UK&30.9X&   1.79X & 3.81X& \best{1.68X}\\
\cline{2-6}
&FS&121.6X &\best{12.44X} &25.23X & 12.62X\\
\hline
\multirow{5}*{SSSP}
&SK&44.3X&  4.23X&3.29X&\best{3.25X}\\
\cline{2-6}
&TW&11.2X& 2.07X& 1.74X& \best{1.25}X  \\ 
\cline{2-6}  
&FK&28.1X &   \best{3.32}X&4.81X&4.60X \\ 
\cline{2-6}
&UK&24.3X &  1.78X&\best{1.11X}&1.13X\\
\cline{2-6}
&FS&24.1X &  3.19X&2.69X&\best{2.52X}\\
\hline
\end{tabular}
\end{table}

We analyze the effectiveness of \sysname's transfer reduction by comparing it with \ptf, Subway (\ptc), and EMOGI (\umzc). We run PageRank and SSSP on all the five real-world graphs and normalize the data transfer volume to the times of edge volume. Table \ref{tab:reduction} shows the results, \ptf has the highest transfer volume. With the help of fine-grained zero-copy access, EMOGI achieves considerable transfer reduction. However, due to the lack of asynchronous scheduling, its transfer volume is still large. {Benefiting from the CPU-based compaction, Subway is expected to have minimal data transfer volume. But the multi-round asynchronous processing performs differently on different algorithms. For PageRank algorithm based on value accumulation, the multi-round processing significantly reduces the transfer times because the additional computations on partitioned subgraphs can still contribute to the final convergence. As processes the transferred subgraph only once more, \sysname has no transfer advantages over Subway for PageRank algorithm, especially on the small graph with few partitions, e.g., \sysname requires 2X data transfer compared to Subway on TW graph. \sysname achieves comparable data transfer volume with subway on SK graph (another small graph) because it benefits a lot from the contribution-driven priority scheduling (As illustrated in Figure 8 (a), the contribution-driven scheduling shows significant performance improvement on the two web graphs, SK and UK.). For the value-replacement-based SSSP algorithm, simply processing the transferred subgraph multiple times may cause stale computation problem (Section \ref{sec:sys:scheduling}), leading to more computations and data transfers. We can observe that Subway transfers more data than EMOGI on SK, TW, UK, and FS for SSSP algorithm. In contrast, with the help of hybrid transfer management and asynchronous task scheduling, \sysname achieves significant transfer reduction in all cases and alleviates the stale computation problem.}

\subsection{Performance Gain of Task Combining and Contribution-Driven Scheduling} 
To analyze the performance gain of task combining and contribution-driven scheduling, we start from the pure hybrid transfer management with basic optimization (multi-stream scheduling) and integrate task combining (as described in section \ref{sec:sys:combine}), and contribution-driven scheduling (as described in section \ref{sec:sys:hub}) one by one. Figure \ref{fig:gain} shows the normalized speedups. The task combining (TC) can bring Hybrid an on average 1.28X, 1.37X, 1.19X, and 1.05X speedups on PageRank, SSSP, CC, and BFS, respectively. The contribution-driven scheduling (CDS) can further bring 2.18X, 1.21X, 1.25X, and 1.06X speedups over the hybrid processing with TC. 
Finally, the two proposed designs can bring an overall 2.78X, 1.67X, 1.47X, and 1.16X speedups over the raw hybrid transfer management, respectively. PageRank algorithm benefits most because the proposed asynchronous processing can effectively accelerate the convergence by prioritizing the vertices with large rank values. In contrast, BFS rarely benefits from the two designs because the vertices are activated only once during the iterative processing.

\begin{figure}
	\centering
	\vspace{-0.05in}
	\includegraphics[width=0.48\textwidth]{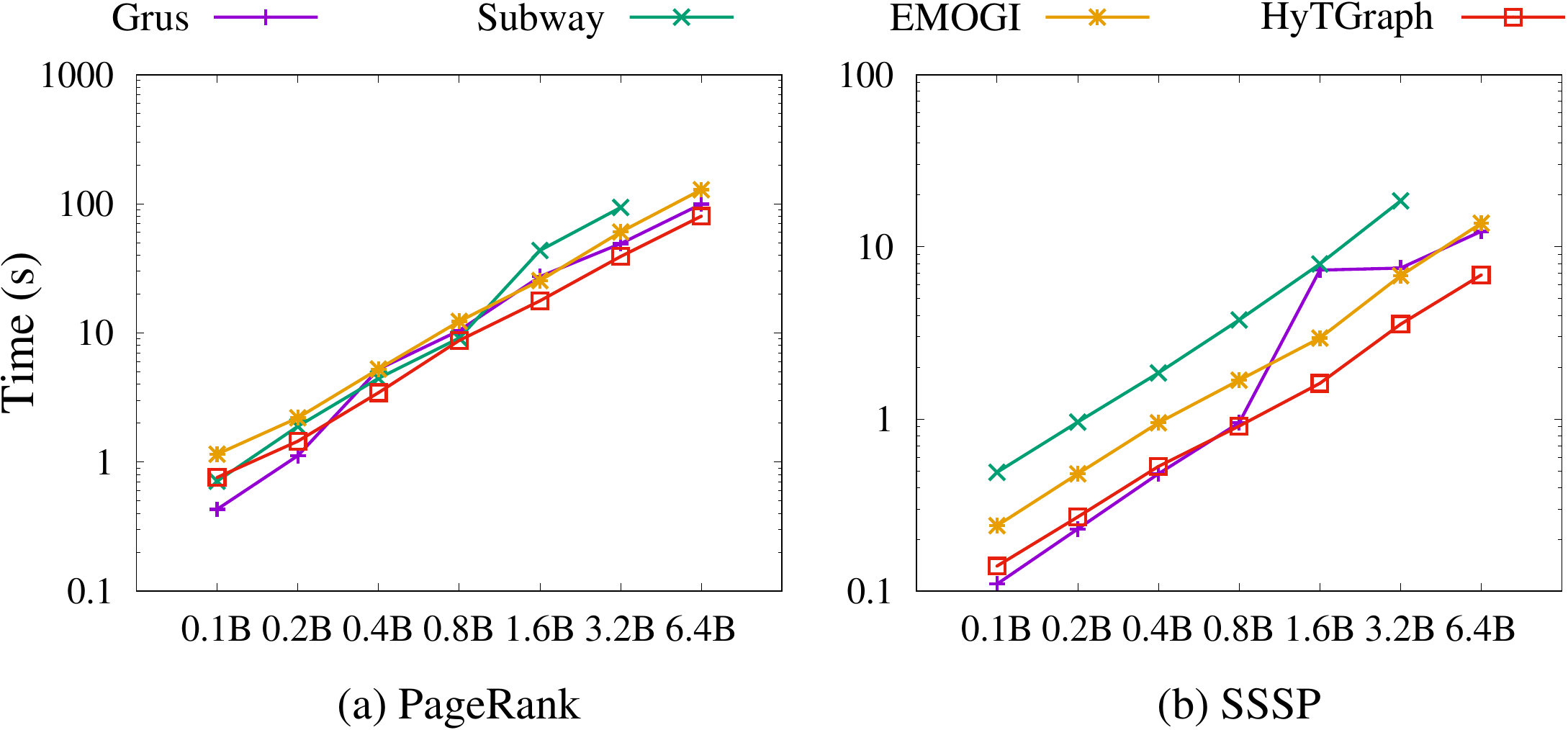}
	\caption{Performance comparison with increasing graph size, the graphs are generated by RMAT with sizes from 0.1 Billion to 6.4 Billion (64X).}
	\label{fig:scale}
	\vspace{-0.1in}
\end{figure}

\subsection{Sensitivity Analysis}
\vspace{-0.05in}
\Paragraph{Varying graph sizes} We compare \sysname with Grus, Subway, and EMOGI under variable graph sizes and report the results in Figure \ref{fig:scale}. When the graph size is small, Grus shows better performance because the data only needs to be loaded once. While, as the graph size increases, the inefficient data transfer of unified-memory will reduce its performance. Subway fails to run the graph with 6.4B edges because of the integer overflow problem. As the graph size increases from 0.1B to 6.4B (64X), the runtimes of Grus, EMOGI, and \sysname for PageRank increase by 231.2X, 111.6X, and 105.39X, respectively. For SSSP algorithm, the runtime of Grus, EMOGI, and \sysname increase by 111.8X, 57.08X, and 49X, respectively. \sysname shows better performance when scaling to larger graphs.

\begin{figure}
	\centering
	\includegraphics[width=0.45\textwidth]{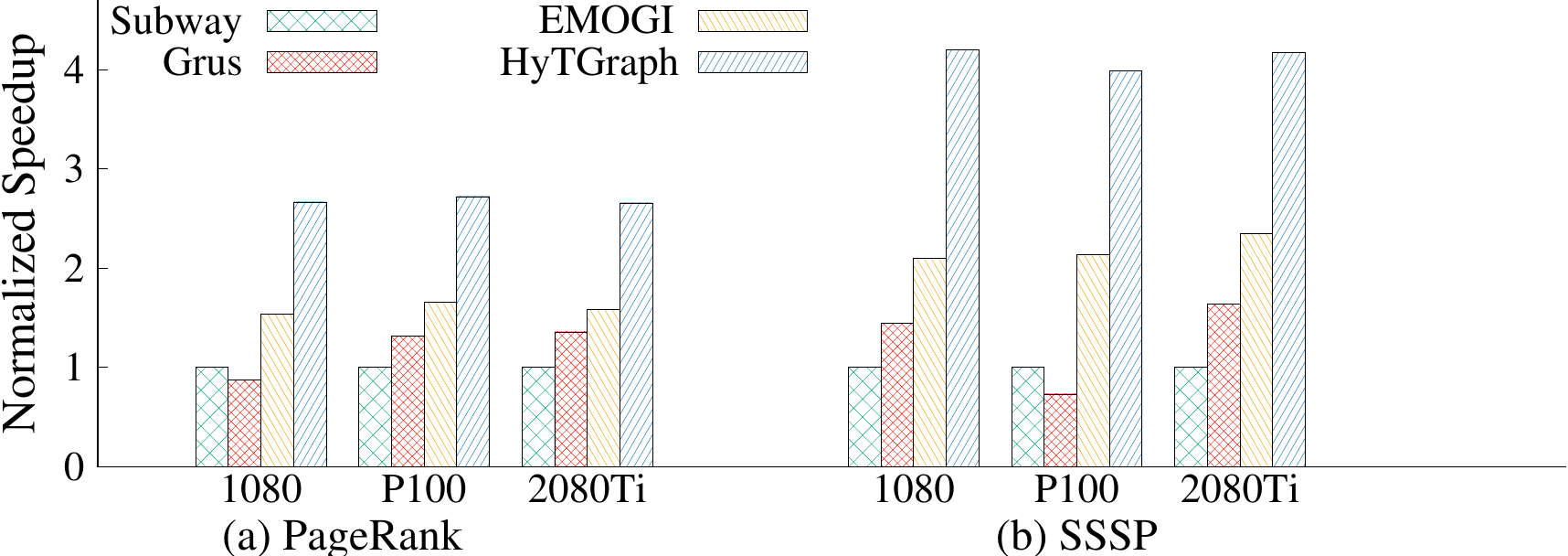}
	\caption{Performance comparison on different GPUs (FS).}
	\label{fig:gpu}
	\vspace{-0.18in}
\end{figure}
\Paragraph{Varying GPUs} We evaluate the performance of \sysname on different GPUs, including GTX 1080 (2560cores, 8GB), TESLA P100 (3584cores, 16GB), and GTX 2080Ti (4352cores, 11GB) with FS graph. We normalize the runtimes of all systems to Subway and show the results in Figure \ref{fig:gpu}. We can observe that \sysname outperforms the other three competitors. For PageRank, \sysname achieves 2.6X-2.7X, 2.0X-3.1X, and 1.6-1.7X speedups over Subway, Grus, and EMOGI, respectively. For SSSP, \sysname achieves 4.0X-4.2X, 2.5X-5.5X, and 1.7X-2.0X speedups over Subway, Grus, and EMOGI, respectively. 

{
\section{Limitations and Future Work}
\vspace{-0.05in}

\Paragraph{Cost computation of \pt-C}
The current version of \sysname uses an approximate method to compute the cost of \pt-C because the overhead of irregular main memory access is hard to quantify accurately. It would be interesting future work to model the \pt-C overhead through machine learning techniques.

\Paragraph{Processing hyper-scale graph}
For a hyper-scale graph whose vertex data exceeds a single GPU memory, processing it with GPU needs to partition the vertex data into smaller chunks that can fit into GPU memory. Such an approach requires frequent host-GPU vertex data swapping, leading to additional data transfer overhead. Therefore, designing new algorithms to optimize the host-GPU vertex data access and exploring whether the computation improvement can cover the additional I/O overhead are interesting and less studied problems. We will take them as our future work.

\Paragraph{Adapting to GPU platforms with fast interconnects} Recently, the hardware makers have come up with fast interconnect technologies (e.g., NVIDIA NVlink\cite{H100} and Intel CXL\cite{CXL}) to replace the slow PCIe bus, which can provide up to 900GB/s GPU-CPU interconnect bandwidth (NVlink-4.0\cite{H100}). In a GPU-accelerated platform with fast interconnections, the main memory may become a new bottleneck of host-GPU data transfer \cite{PUMP_SIGMOD_2020}. We can improve \sysname by exploring the main memory access performances of different transfer methods and integrating the main memory accessing cost in our hybrid model to adapt to these new platforms.
}

%% file: sec6.tex
\vspace{-0.05in}
\section{Related Work}
\vspace{-0.05in}
\label{sec:related}

\Paragraph{In-GPU-memory graph processing}
To accelerate graph processing, the high parallelism of GPU has attracted great attention\cite{harish_hipc_2007,xbfs_hpdc_2019,cuda_ppopp_2011,cusha_hpdc_2014,duane_acmtrans_2015,gunrock_ppopp_2016,digraph_asplos_2019,medusa_tpds_2014}. Cusha\cite{cusha_hpdc_2014} uses two novel data structures, named GShards and CW, to avoid non-coalesced memory access. Gunrock\cite{gunrock_ppopp_2016} performs computation on the frontier with data-centric abstraction. Tigr\cite{tigr_asplos_2018} proposes a virtual transformation to transform skewed graphs into virtual vertices for load-balancing. SEP-Graph\cite{sep-graph_ppopp_2019} switches execution paths adaptively based on a selection in each of the three pairs of parameters, namely, Sync or Async, Push or Pull, and DD (data-driven) or TD (topology-driven). 

\Paragraph{Out-of-GPU-memory graph processing}
GPU-accelerated graph processing has attracted extensive attention. Besides the systems mentioned above\cite{emogi_vldb_2020,halo_vldb_2020,grus_acmtrans_2021,graphie_pact_2017,graphreduce_sc_2015,scaph_atc_2020,subway_eurosys_2020,ASCETIC_ICPP_2021}, recent studies also propose CPU-GPU co-processing to accelerate large graphs computation \cite{totem_pact_2012,garaph_atc_2017}. Totem \cite{totem_pact_2012} partitions a graph into two subgraphs, one for the CPU and one for GPU, keeping the number of data transfers to a minimum at the expense of severe load imbalance. Garaph \cite{garaph_atc_2017} concurrently processes the active subgraphs on both the host and GPU. However, the CPU-based low-activeness subgraph processing may become a new bottleneck.

\section{Conclusion}
\vspace{-0.05in}
We present \sysname, a highly efficient GPU-accelerated graph processing framework by adaptively switching the transfer management approach involving explicit transfer management and implicit transfer management. This hybrid approach maximizes the host-GPU bandwidth and is necessary to achieve the shortest overall execution time. Our intensive experiments show the high effectiveness of \sysname.